\documentclass[aps,prd,twocolumn,nofootinbib,superscriptaddress]{revtex4}
\usepackage[hypertex]{hyperref}
\usepackage{graphicx} 
\usepackage{amsbsy}   
\newcommand{\sphi}{S_{\phi K_S}}
\newcommand{\seta}{S_{\eta' K_S}}

\newcommand{\dtlbar}[1]{{\,\overline{\!#1}{}}}
\begin{document}

\pagestyle{plain}

\title{Right-Handed New Physics Remains Strangely Beautiful}

\author{Daniel T. Larson}
\affiliation{Theoretical Physics Group, 
Ernest Orlando Lawrence Berkeley National Laboratory,
University of California, Berkeley, CA 94720}
\affiliation{Department of Physics, University of California,
Berkeley, CA 94720}

\author{Hitoshi Murayama}
\affiliation{Theoretical Physics Group, 
Ernest Orlando Lawrence Berkeley National Laboratory,
University of California, Berkeley, CA 94720}
\affiliation{Department of Physics, University of California,
Berkeley, CA 94720}

\author{Gilad Perez}
\affiliation{Theoretical Physics Group, 
Ernest Orlando Lawrence Berkeley National Laboratory,
University of California, Berkeley, CA 94720}

\date{\today}

\begin{abstract}
Current data on CP violation in $B_d \rightarrow \eta' K_S$ and $B_d
\rightarrow \phi K_S$, taken literally, suggest new physics
contributions in $b\rightarrow s$ transitions.  Despite a claim to the
contrary, we point out that right-handed operators with a single weak
phase can account for both deviations thanks to the two-fold ambiguity
in the extraction of the weak phase from the corresponding
CP-asymmetry. This observation is welcome since large mixing in the
right-handed sector is favored by many GUT models and frameworks which
address the flavor puzzle.  There are also interesting correlations
with the $B_s$ system which provide a way to test this scenario in the
near future.
\end{abstract} \pacs{Who cares?} \maketitle

\section{Introduction}
\label{sec:intro}

Recent years have seen remarkable progress in flavor physics. The
dramatic discovery of neutrino oscillations~\cite{nu-oscillations},
long suspected to be the explanation for the solar neutrino deficit,
is the first firm evidence for physics beyond the Standard Model
(SM). In the quark sector, $T$-violation was discovered in the neutral
kaon system~\cite{Angelopoulos:1998dv} in agreement with observed
indirect CP violation and the CPT theorem. The first example of direct
CP violation was also discovered in the neutral kaon
system~\cite{FirstDir}. Finally, both indirect and direct CP violation
have been observed for the first time in another system, in the decays
$B_d \rightarrow J/\psi K_S$ and $B_d \rightarrow K\pi$
respectively~\cite{psiK,DirCPB}.  The data confirms that the CP
violation observed to date originates in the Kobayashi--Maskawa (KM)
phase of the quark mixing matrix. It is incredible that all of this
has been accomplished in a six-year period! However, the origin of
flavor and patterns of masses and mixings still remain elusive.

Interestingly, a deviation from the KM theory has been reported in
$B_{d}^{0} \rightarrow \phi K_{S}$ and $B_{d}^{0} \rightarrow \eta'
K_{S}$. The time-dependent CP-asymmetries in these modes, denoted
$\sphi$ and $\seta$ respectively, yield an effective value of $\sin
2\beta=\sin 2\phi_1$ that differs from that in the $J/\psi K_S$ final
state.\footnote{The SM predicts this deviation is at most ${\cal
O}(0.1)$. For more details
see~\cite{London:1997zk,Grossman:1996ke,Beneke:2003zv,Grossman:2003qp}.}
The averages of Belle and BaBar~\cite{BaBe} are, according
to~\cite{Ligeti:2004ak}:
\begin{equation}
  \sphi = 0.34\pm 0.21,\ \ \ \seta = 0.41\pm 0.11, 
  \label{exp}
\end{equation}
which should be compared to
\begin{equation}
  S_{J/\psi K_S} = 0.726\pm 0.037 .
\end{equation}
The discrepancy is $2.7\sigma$ for the $\eta' K_S$ mode, and the two
modes show values consistent with each other.\footnote{There is a
slight inconsistency in the measurements of $\seta$ at BaBar and
Belle~\cite{BaBe}.}

The evidence for a new source of CP violation is not yet conclusive.
Nonetheless, many new physics models have been shown to contribute
significantly to the CP violation in $b \rightarrow s$ transitions
within the limits from other experimental constraints.  For example,
the observed large mixing between $\nu_\mu$ and $\nu_\tau$, once
grand-unified, can lead to a large mixing between $\tilde{s}_R$ and
$\tilde{b}_R$~\cite{SO10,Harnik:2002vs}. There is a growing
excitement to see if such new physics scenarios can account for the
observed difference between $\sphi$, $\seta$ and $S_{J/\psi K_S}$.

The fact that $\sphi$ and $\seta$ are similar yields information about
the operators responsible for the deviation from $S_{J/\psi K_S}$, and
hence a hint as to the underlying new physics. Contributions to $B$
decays proceed through effective operators of two types,
$\mathcal{O}_i$ and $\widetilde{\mathcal{O}}_i$ (see {\it
e.g.}~\cite{Harnik:2002vs} for definitions of the relevant
operators). The Standard Model contributes only to ``left-handed''
(LH) operators $\mathcal{O}_i$, while some new physics scenarios, such
as mixing between right-handed squarks, only contribute to the
``right-handed'' (RH) operators $\widetilde{\mathcal{O}}_i$. Several
groups~\cite{Khalil:2003bi,Kagan:2004ia,Endo:2004dc} have made the
observation that $\langle f|\mathcal{O}_i|B\rangle = (-1)^{P_f+1}
\langle f|\widetilde{\mathcal{O}}_i|B\rangle$ where $P_f$ is the
parity of the final state. Since $\phi K$ is parity odd (a
pseudoscalar and vector in a $p$-wave state) while $\eta' K$ is parity
even (two pseudoscalars in an $s$-wave state), these two final states
will be sensitive to different combinations of the $\mathcal{O}_i$ and
$\widetilde{\mathcal{O}}_i$ operators.

In~\cite{Endo:2004dc} it is claimed that the recent measurements of
the CP-asymmetries $\sphi$ and $\seta$ imply that the additional
CP-violating phases appear in the mixing of left-handed squarks and
not right-handed squarks. Here we demonstrate that this is not
necessarily the case.  Because of the two-fold ambiguity in the
extraction of the weak phase from $\sphi$ and $\seta$,
right-handed operators with a single new CP phase can successfully
account for the observed discrepancies.  The preferred parameter
region is more tightly constrained than that for the opposite
chirality, but this allows for a more precise study of correlations
with other observables such as $B_s$ mixing and $S_{\psi\phi}$. Thus
we may be able to test this scenario in the near future.

\section{Simplified General Analysis}
\label{sec:simple}

We are interested in a scenario in which the new physics contribution
is dominated by right-handed operators with a single source of CP
violation. In Section~\ref{sec:discussion} we will comment on the role
of strong phases, but for simplicity we shall begin by assuming that
all strong phases are negligible. In that case the amplitude for the
$B$-meson decay can be written
\begin{equation}
{\mathcal A}(B^0\rightarrow \phi,\eta' )={\mathcal
A}^\mathrm{SM}_{\phi,\eta'}\left( 1\pm r_{\phi,\eta'}
e^{i\sigma_s}\right)
\label{Amp}
\end{equation}
where $\sigma_s$ is the weak phase of the RH operators and
\begin{equation}
  r_{\phi,\eta'}\equiv \left|
  \frac{\mathcal{A}^\mathrm{NP}_{\phi,\eta'}}
  {\mathcal{A}^\mathrm{SM}_{\phi,\eta'}} 
  \right| \,.
\label{AMI}
\end{equation}
It is known that the matrix elements for $B$ decays normalized to the
SM contributions are similar for the final states $\phi K$ and $\eta'
K$~\cite{Khalil:2003bi}, so as a first approximation we will assume
that they are in fact identical. This means
\begin{equation}
r_{\phi}=r_{\eta'}\equiv r\,.\label{rassum}
\end{equation}
We will use this approximation for our general analysis, but will come
back to discuss deviations from this assumption in
Section~\ref{sec:realistic}.

The mixing induced CP asymmetry is given by
\begin{equation}
S_{\phi,\eta' } = \mathrm{Im} \left(-\frac{\bar A_{ \phi,\eta'}}{A_{
\phi,\eta'}}
\frac{V_{td}V_{tb}^*}{V_{td}^*V_{tb}}
\frac{V_{cs}V_{cd}^*}{V_{cs}^*V_{cd}} \right),
\label{SMI} 
\end{equation}
where the second factor is the $B^0-\dtlbar{B^0}$ mixing, and the third
factor is the $K^0-\dtlbar{K^0}$ mixing. Using Equation~(\ref{Amp})
and assuming the SM contribution is purely real we find
\begin{equation}
S_{\phi,\eta' } = \frac{\sin(2\beta)\pm2r \sin(2\beta+\sigma_s)+
r^2\sin(2\beta+2\sigma_s)}{1\pm2r \cos \sigma_s +r^2} 
\label{SMI1}
\end{equation}
where $V_{td}=\left|V_{td} \right| e^{-i\beta}\,.$

It is useful to first consider the two interesting limits of small and
large $r$.  For small $r$ we find
\begin{equation}
  \lim_{r\to0}\left(S_{\phi,\eta' }\right)\approx \sin(2\beta)\pm 2r
  \cos2\beta
  \sin\sigma_s
  \label{SMI1s}\,.
\end{equation}
Since the measured $\sphi$ and $\seta$ are both smaller than $\sin
2\beta$, it is clear that a single RH operator cannot account
for the data when $r$ is small. On the other hand, when $r$ is large we find
\begin{equation}
  \lim_{r\to\infty}\left(S_{\phi,\eta' }\right)\approx
  \sin(2\beta+2\sigma_s)\mp {2\over r}\cos(2\beta+2\sigma_s) \sin\sigma_s 
  \label{SMI1l}\,,
\end{equation}
which naturally leads to deviations for $\sphi$ and $\seta$ in the
same direction. In reality we expect that $r$ will lie somewhere
between these two extremes. But studying these limits shows that RH
operators require sizeable new physics contributions ($r\gtrsim
\mathcal{O}(1)$) to account for the data. This contrasts with the case
of LH operators where new physics contributions can be as small as $r\sim 0.2$.

\begin{figure}[tc]
\centering \includegraphics[width=\columnwidth]{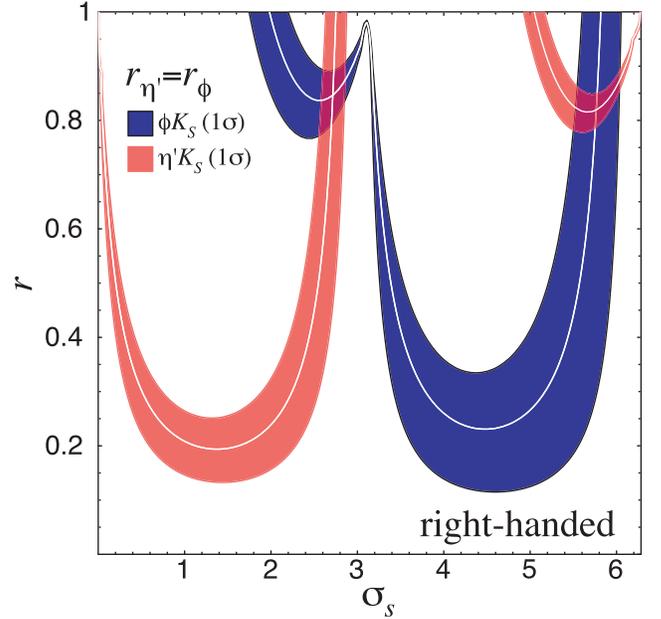}
\caption{Central value and 1-sigma contours of $\sphi$ (blue) and
$\seta$ (pink) for a RH new physics contribution as a function of
$\sigma_s$ and $r=r_\phi=r_{\eta'}$.}
\label{fig:rhcontour}
\end{figure}

We can analyze the situation more precisely by plotting $\sphi$ and
$\seta$ as functions of $r$ and $\sigma_s$. Figure~\ref{fig:rhcontour}
shows the central value and one-sigma contours of $\sphi$ and
$\seta$. The bands overlap in the two regions around $(\sigma_s,
r)\sim(2.7,0.8)$ and $(5.7,0.8)$. This implies that a single RH
operator \emph{can} account for the experimental results, despite the
claim in~\cite{Endo:2004dc}. In addition, since there are only two
discrete regions that are consistent with the data, we can hope to
find strong correlations with other observables.

\begin{figure}[tc]
\centering \includegraphics[width=\columnwidth]{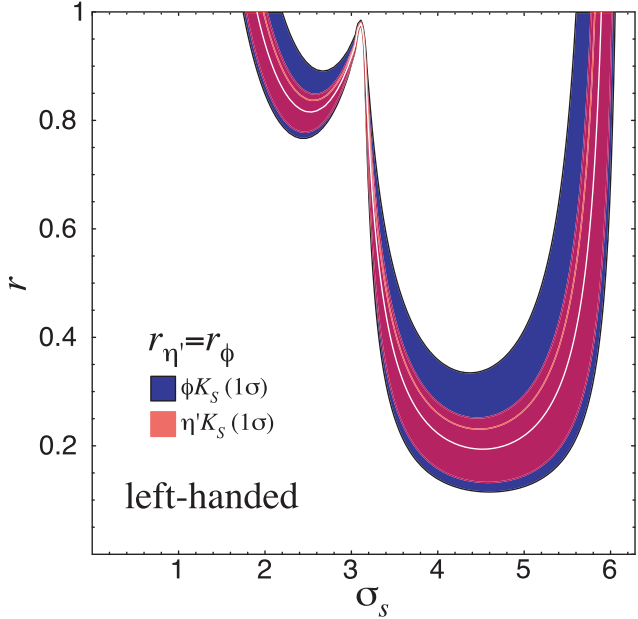}
\caption{Central value and 1-sigma contours of $\sphi$ (blue) and
$\seta$ (pink) for a LH new physics contribution as a function of
$\sigma_s$ and $r=r_\phi=r_{\eta'}$.}
\label{fig:lhcontour}
\end{figure}

\begin{figure}[tc]
\centering \includegraphics[width=\columnwidth]{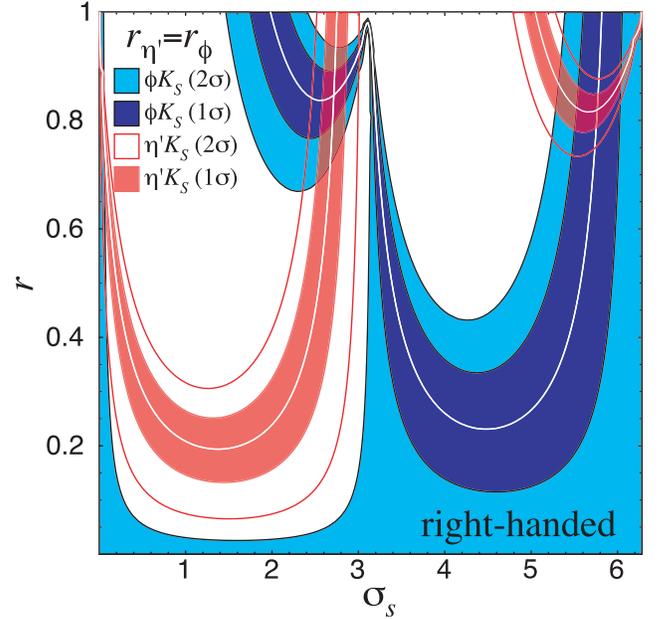}
\caption{Central value, 1- and 2-sigma contours of $\sphi$ (blue) and
$\seta$ (pink) for a RH new physics contribution as a function of
$\sigma_s$ and $r=r_\phi=r_{\eta'}$.}
\label{fig:2sigma}
\end{figure}

For comparison, in Figure~\ref{fig:lhcontour} we show the same
contours but this time assuming the new physics contribution comes
from LH operators. As expected, since there is no longer a
relative sign difference between the new physics contributions to
$\sphi$ and $\seta$, the overlap regions are much larger. Thus a wider
range of parameters is consistent with the data, but that also means less
information can be extracted. The difference between
Figures~\ref{fig:rhcontour} and~\ref{fig:lhcontour} is what led to the
the conclusion in~\cite{Endo:2004dc} that new physics contributions to
LH operators are favored, but we clearly see that contributions from
RH operators are not excluded yet, especially if one considers the
$2\sigma$ allowed regions, which are shown in Figure~\ref{fig:2sigma}.

These remaining allowed regions for RH operators are important because
flavor violation from new degrees of freedom is aligned with that of
the SM in many simple flavor models such as Abelian~\cite{Abel} and
non-Abelian~\cite{Ross} horizontal models, split fermion
models~\cite{split} and RS1 models~\cite{RS1}.  This implies the
following relation between the quark masses and left and right
diagonalization matrices $D_{R,L}$ of the down type Yukawa matrix:
\begin{equation}m_s/m_b\sim \left(D_L\right)_{23}
\left(D_R\right)_{23}\lesssim  \left(V_{\rm CKM}\right)_{23}
\left(D_R\right)_{23}\,,\end{equation}
which implies
\begin{equation}
\left(D_R\right)_{23}={\cal O}(1)\,.
\end{equation}
Thus we generically expect new physics to induce RH operators which
are not suppressed by $V_{cb}$ and therefore might yield the dominant
contribution. Also, constraints from some measurements such as
$b\rightarrow s\gamma$ are weaker for RH operators because the LH
contributions add coherently to the SM whereas RH contributions add
incoherently.

The reason our conclusion differs from the one presented
in~\cite{Endo:2004dc} is that we rely on the discrete ambiguities in
the sine function.  In the
presence of new physics contributions we can parameterize the $CP$
asymmetries as
\begin{equation}
S_{\phi,\eta'}=\sin(2\beta+\Sigma_{\phi,\eta'})
\label{eqn:sigma}
\end{equation}
where
\begin{equation}
\Sigma_{\phi,\eta'}=\arg\left({1\pm r e^{i\sigma_s}\over 1\pm r
e^{-i\sigma_s}}\right)
\,
\end{equation}
is the phase coming from the decay amplitude. There are
two solutions to Equation~(\ref{eqn:sigma}), namely
\begin{eqnarray}
\Sigma_{\phi,\eta'}&=&
{\rm arcsin}\left(S_{\phi,\eta'}\right)-2\beta \label{us}\\
&\sim& \frac{\pi}{8}-\frac{\pi}{4}
 = -\frac{\pi}{8} \sim -0.39
\end{eqnarray}
and
\begin{eqnarray}
\Sigma_{\phi,\eta'} &=& \pi-2\beta-\mathrm{arcsin}
\left(S_{\phi,\eta'}\right) \\
&\sim& \pi - \frac{\pi}{4}-\frac{\pi}{8}
= \frac{5\pi}{8}\sim 1.96\, ,
\end{eqnarray}
where in the second line of each solution we have used the rough
approximation $\sphi\sim\seta\sim \sin(\pi/8)$ and $2\beta \sim
\pi/4$.  As argued above, for a RH operator to give a contribution
such that $\Sigma_\phi \sim \Sigma_{\eta'}$ we would need $r\gg 1$,
which would likely already have been observed. Thus we are led to
consider a mixed scenario, which means we expect solutions centered
around $(\Sigma_\phi, \Sigma_{\eta'}) = (-\pi/8, 5\pi/8)$ or $(5\pi/8,
-\pi/8)$.  This agrees with the numbers one extracts from the regions of
overlap in Fig.~\ref{fig:rhcontour}.

\section{More Realistic Analysis}
\label{sec:realistic}

In the previous section we have analyzed the situation where the size
of the new physics amplitude relative to the SM is the same for both
the $\phi K$ and $\eta' K$ final states. This assumption can be
checked using a specific hadronic model. Within the framework of naive
factorization we found that $r_\phi/ r_\eta' \sim \mathcal{O}(1)$ for the
relevant operators. For instance the chromomagnetic operator,
which often gives the dominant contribution, yields $r_\phi/
r_\eta'\simeq0.9$, demonstrating that this assumption is reasonable.

\begin{figure}[tc]
\centering \includegraphics[width=\columnwidth]{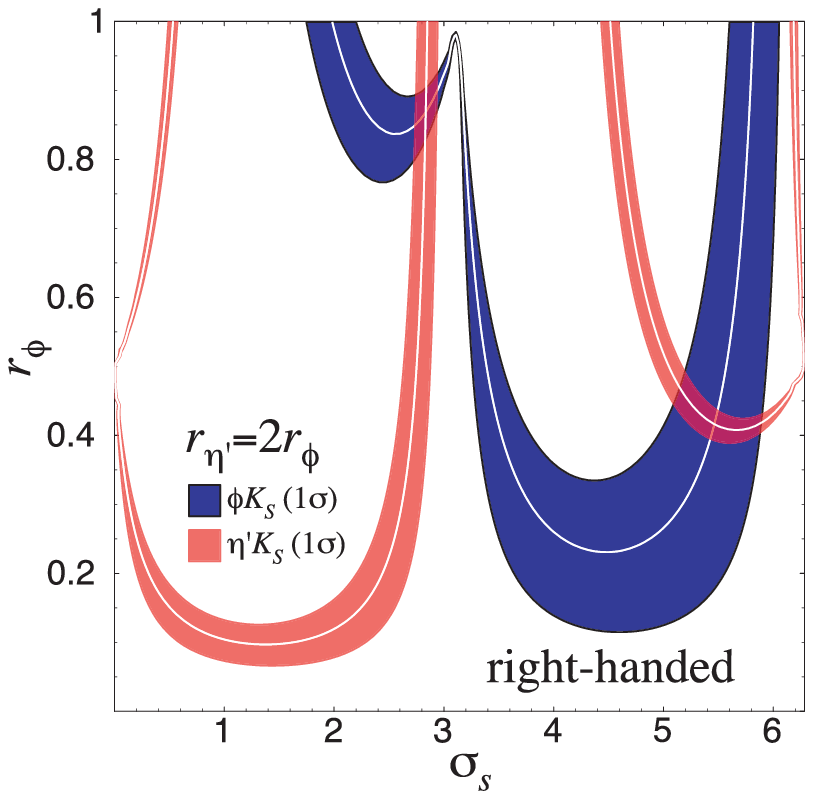}
\caption{Central value and 1-sigma contours of $\sphi$ (blue) and
$\seta$ (pink) for a RH new physics contribution as a function of
$\sigma_s$ and $r_\phi$, with $r_{\eta'} = 2 r_\phi$.}
\label{fig:rhcontour2}
\end{figure}

\begin{figure}[tc]
\centering \includegraphics[width=\columnwidth]{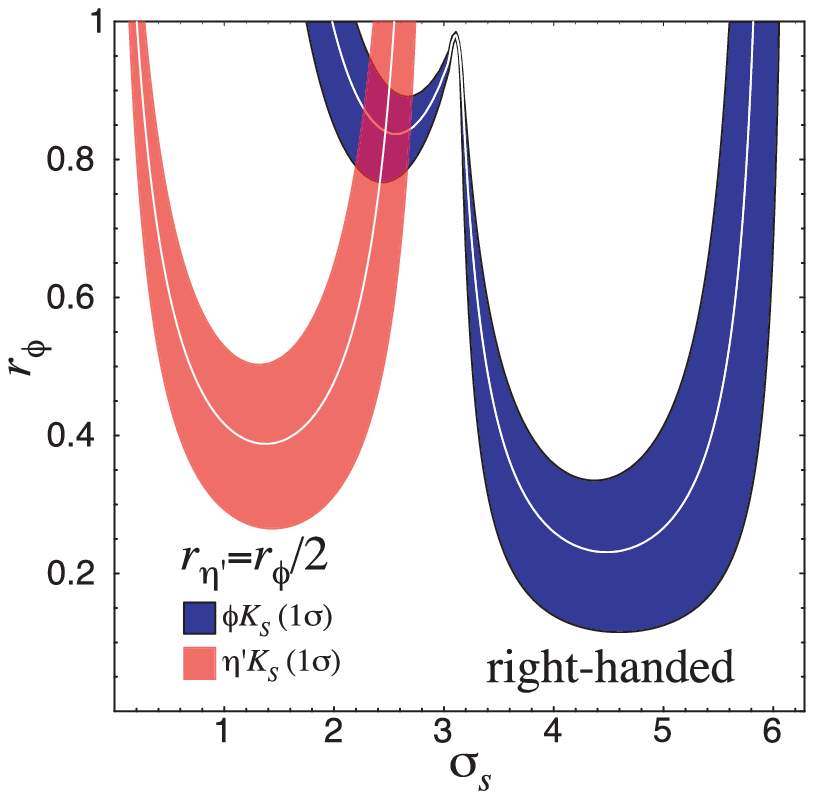}
\caption{Central value and 1-sigma contours of $\sphi$ (blue) and
$\seta$ (pink) for a RH new physics contribution as a function of
$\sigma_s$ and $r_\phi$, with $r_{\eta'} = r_\phi/2$.}
\label{fig:rhcontourhalf}
\end{figure}

\begin{figure}[tc]
\centering \includegraphics[width=\columnwidth]{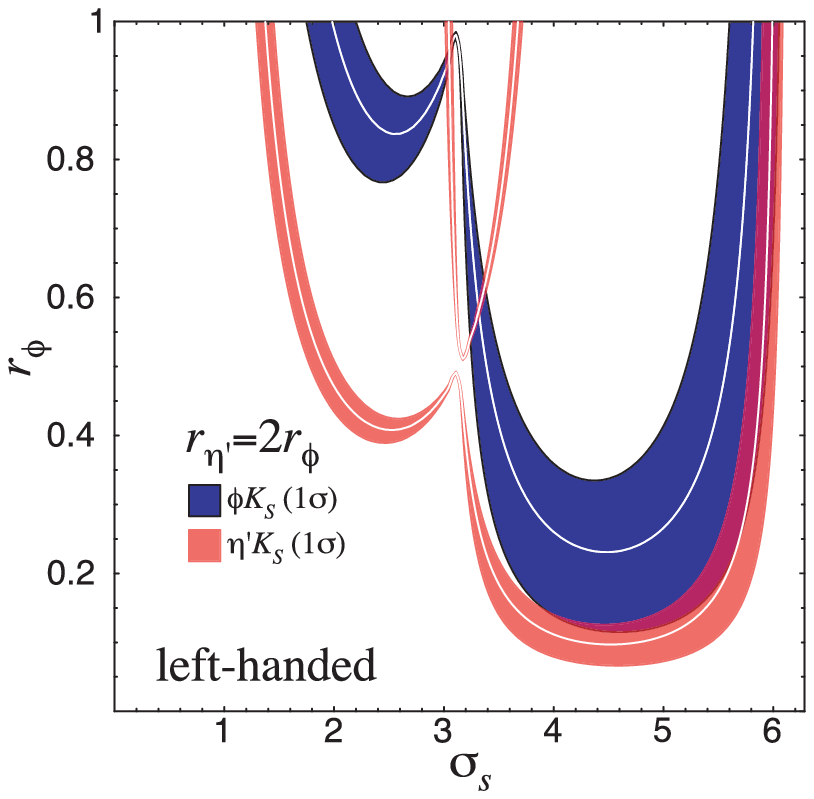}
\caption{Central value and 1-sigma contours of $\sphi$ (blue) and
$\seta$ (pink) for a LH new physics contribution as a function of
$\sigma_s$ and $r_\phi$, with $r_{\eta'} = 2 r_\phi$.}
\label{fig:lhcontour2}
\end{figure}

\begin{figure}[tc]
\centering \includegraphics[width=\columnwidth]{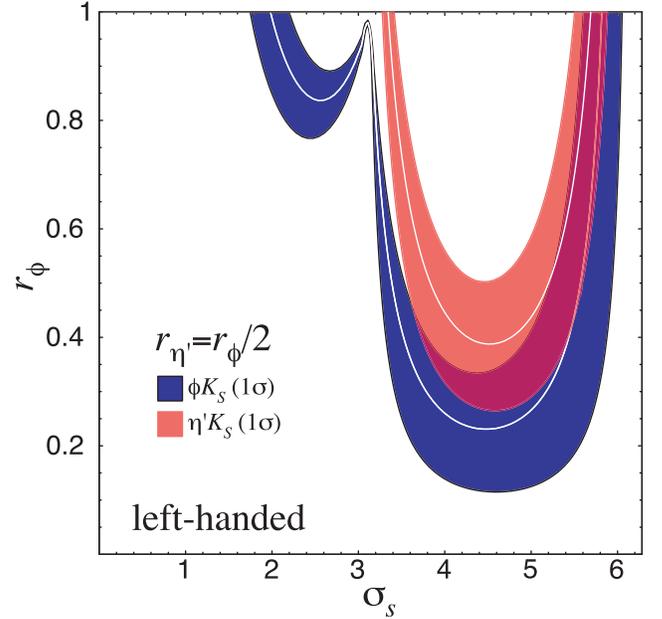}
\caption{Central value and 1-sigma contours of $\sphi$ (blue) and
$\seta$ (red) for a LH new physics contribution as a function of
$\sigma_s$ and $r_\phi$, where $r_{\eta'} = r_\phi/2$.}
\label{fig:lhcontourhalf}
\end{figure}

We can relax the assumption in Equation~(\ref{rassum}) but still
assume that the new physics is governed by a single source of CP
violation.  In this situation we expect order one variations between
$r_\phi$ and $r_{\eta'}$.\footnote{Applying the analysis presented in
\cite{Harnik:2002vs} to estimate the hadronic matrix elements, we have
checked that for a specific model with RH squark mixing between the
2nd and 3rd generations $r_\phi$ and $r_{\eta'}$ are within a factor
of 2 over most of the parameter space.} To demonstrate the effect of
variations of $r$, we plot contours of $\sphi$ and $\seta$ as a
function of $\sigma_s$ and $r_\phi$ with $r_{\eta'}=2r_\mathrm{\phi}$
(Figure~\ref{fig:rhcontour2}) and $r_{\eta'}=r_\mathrm{\phi}/2$
(Figure~\ref{fig:rhcontourhalf}). These figures show that, despite the
shift in the contours, regions of overlap still remain where RH
operators can account for the deviations in $\sphi$ and $\seta$ from
$S_{J/\psi K_S}$, even when $r_\phi \neq r_{\eta'}$.  In fact, for LH
operators the overlap of the contours starts to separate when $r_\phi
\neq r_{\eta'}$ as shown in Figures~\ref{fig:lhcontour2}
and~\ref{fig:lhcontourhalf}. This demonstrates that the ease with
which the LH operators can fit the data is dependent on the similarity
between the relative size of new physics contributions to $\phi K$ and
$\eta' K$ final states.

In reality there are further complications. The presence of strong
phases that differ for $\phi K$ and $\eta' K$ final states will
generally lead to different dependence on $\sigma_s$, reducing the
strong correlations between the two modes. Also, the absorptive part
of the $c\bar{c}$ loop needs to be taken into consideration, which
introduces another strong phase. The situation is less clear-cut if
there are multiple new CP phases in the new physics sector. Finally,
the uncertainties in hadronic matrix elements will always add further
complication. Eventually all of these subtleties must be addressed,
but at this early stage of the analysis it is important to study the
general features and see how far that will take us.

\section{Correlations with the $B_s-\dtlbar{B}_s$ system}

\begin{figure}[tc]
\centering \includegraphics[width=\columnwidth]{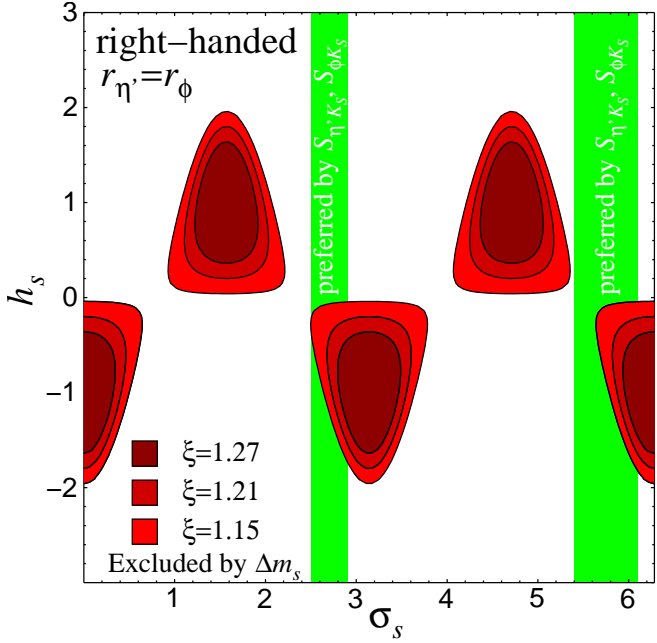}
\caption{The $\sigma_s-h_s$ plane showing the regions excluded by the
lower bound on $\Delta m_s$ (red ovals) and also the regions
preferred by $\sphi$ and $\seta$ as given in
Equation~(\ref{sigsbound}) (green bands).}
\label{fig:dmsphi}
\end{figure}

Let us continue with our assumption that new physics contributions are
dominated by a single source of CP violation which appears only in the
RH operators.  This implies that we can parameterize the contribution
to the $B_s-\dtlbar{B}_s$ transition amplitude, $M^s_{12}$, as
\begin{equation}
M_{12}^s={M_{12}^s}^{\rm SM}\left(1+h_s e^{2i\sigma_s}\right)\,,
\label{M12s}
\end{equation}
where $h_s$ is the ratio of new physics to SM contributions, and
$\sigma_s$ is the same phase that appears in Equation~(\ref{Amp}) for
the $B_d$ decays. In principle $h_s$ could be quite large, though we
expect it to be order one in most models. Note also that the sign of
$h_s$ is model dependent, so in our analysis below we consider both
positive and negative values. With this parameterization the mass
difference between $B_s^0$ and $\dtlbar{B^0}_s$ is given by
\begin{equation}
\Delta m_s=\Delta m_s^{\rm SM} \left|1+h_s e^{2i\sigma_s} \right|.
\label{dmsNP}
\end{equation}
So far experiments have only yielded a lower bound on the value of the
mass difference, $\Delta m_s^{\rm exp} > 14.5\mbox{ ps}^{-1}$~\cite{hfag}.
The SM prediction of $\Delta m_s$ has significant hadronic
uncertainties, but a much cleaner prediction is available for the ratio
$\Delta m_s/\Delta m_d\,.$ Since the new physics
contribution to $b\to d$ transitions is very tightly constrained by
current measurements~\cite{Ligeti:2004ak}, we will assume
that $\Delta m_d^{\rm}\sim \Delta m_d^{\rm exp}\,.$ Consequently we
can use the state of the art analysis~\cite{CKMfitter} and
Equation~(\ref{dmsNP}) to find the following relation between the predicted
ratio for $\Delta m^\mathrm{SM}_s/\Delta m^\mathrm{SM}_d$ and the
experimental lower bound on
$\Delta m_s^{\rm exp}/\Delta m_d^{\rm exp}$ 
\begin{eqnarray}
\left|1+h_s e^{2i\sigma_s} \right| &=& {\Delta m_d^{\rm
SM}\over\Delta m_s^{\rm SM}}\,{\Delta m_s^{\rm exp}\over\Delta
m_d^{\rm exp}} =  \frac{m_{B_d}}{m_{B_s}} \frac{1}{\xi^2}
\frac{|V_{td}|^2}{|V_{ts}|^2} \nonumber\\
&\gtrsim&0.8 \,, 
\label{delmscon} 
\end{eqnarray} 
with a 20\% uncertainty coming from the CKM factors. Here
$\xi=f_{B_s}/f_{B_d}\,\sqrt{B_{B_s}/B_{B_d}}\sim1.21\pm0.06
$~\cite{CKMfitter} is related to the ratio between the bag parameters
and $m_{B_d}/m_{B_s}\approx 0.98$.
Equation~(\ref{delmscon}) provides a constraint on the allowed values
of $h_s$ and $\sigma_s$. The constraint is, at present, rather weak
and corresponds to several oval excluded regions shown in Figure
\ref{fig:dmsphi}. Examining the regions of intersection for the
1-sigma contours in Figure~\ref{fig:rhcontour}, we find the following
range for $\sigma_s$ preferred by $\sphi$ and $\seta$:
\begin{equation}
\sigma_s=2.5-2.9\, \ {\rm or} \, \ 5.4-6.1\,.\label{sigsbound}  
\end{equation}
This corresponds to the two vertical strips shown in
Figure~\ref{fig:dmsphi}. At present the regions excluded by the lower
bound on $\Delta m_s$ partially intersect with the regions preferred
by $\sphi$ and $\seta$ only for negative values of $h_s$. However, in
the near future measurements at the Tevatron, LHC-b, and BTev will
either measure $\Delta m_s$ or significantly raise the lower bound, so
the above constraint will play an important role in discriminating
between new physics frameworks.

\begin{figure}[tc]
\centering \includegraphics[width=\columnwidth]{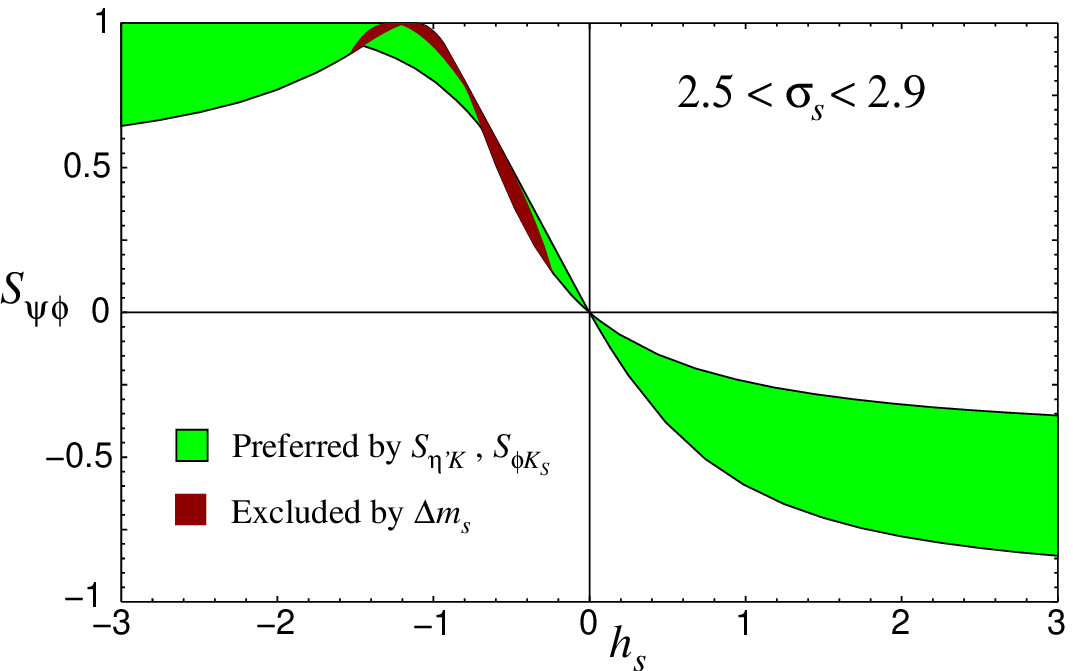}
\centering \includegraphics[width=\columnwidth]{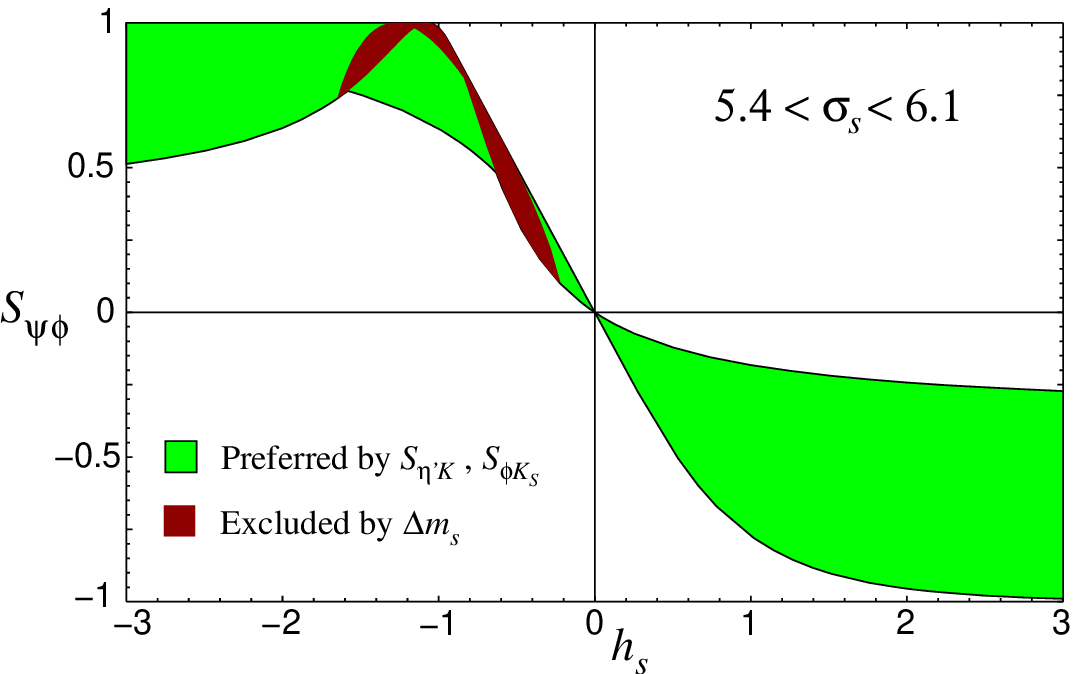}
\caption{Allowed regions for $S_{\psi\phi}$ as a function of $h_s$
(green) and the regions excluded by the lower bound on $\Delta
m_s$ (red). The upper plot corresponds to $\sigma_s \in [2.5,
2.9]$ while the lower plot corresponds to $\sigma_s \in
[5.4,6.1]$. The ranges for $\sigma_s$ are those compatible with the
measured values of $\sphi$ and $\seta$.}
\label{fig:psiphi}
\end{figure}

The requirements on $\sigma_s$ from $\sphi$ and $\seta$ are correlated
with the prediction for $S_{\psi\phi}$, the CP asymmetry in
$B_s\to\psi\phi$. Within the Standard Model, $S_{\psi\phi}\lesssim
\lambda^2$, so any CP-asymmetry will only arise from the new phase in
the contribution to $B_s$ mixing shown in Equation~(\ref{M12s}). Thus
we find
\begin{eqnarray}
S_{\psi\phi}&=&\sin\left[ {\rm arg}\left(1+h_s
e^{2i\sigma_s}\right)\right] \nonumber\\
&=& \frac{h_s\sin 2\sigma_s}{\sqrt{(1+h_s\cos 2\sigma_s)^2
+ (h_s\sin 2\sigma_s)^2}}\,.
\end{eqnarray}
Figure~\ref{fig:psiphi} shows the allowed values for $S_{\psi\phi}$ as
a function of $h_s$ when $\sigma_s$ is in the range
specified by Equation~(\ref{sigsbound}). In addition, the regions
excluded by the lower bound on $\Delta m_s$ are also shown. Already we
see some very non-trivial correlations between measurements of
$\sphi$, $\seta$ and $S_{\psi\phi}$ that will only get stronger as
the data improve.

\section{Discussion}
\label{sec:discussion}

So far we have assumed that all strong phases are negligible. In the
framework of naive factorization this is expected. However, there is
now some evidence to the contrary due to the observation of direct
$CP$-violation in the decay $\overline{B^0}\rightarrow
K^+\pi^-$~\cite{DirCPB}. The presence of large strong phases, which
should generically be different for $\phi K$ and $\eta' K$ final
states, complicate the analysis and make it more sensitive to the
framework used to estimate the hadronic matrix elements.
 
When the new physics contribution is large one would generically
expect to measure direct $CP$-violation. Currently the experimental
limits are $|C_{\phi K_S}| < 0.2$ and $|C_{\eta' K_S}| <
0.1$~\cite{Ligeti:2004ak}.  Using isospin symmetry one finds much
stronger constraints from measurements of the CP asymmetries, $A^{\rm
CP}_{\phi,\eta' K^\pm}$, in the charged modes $B^\pm\to \phi K^\pm,\,
\eta' K^\pm$.\footnote{Assuming that there is no exact cancellation
between rescattering amplitudes~\cite{Grossman:2003qp}.}  Using the
world averages one finds~\cite{hfag}
\begin{equation}
A^{\rm CP}_{\phi K^\pm}=0.04\pm0.05\,,\ \ A^{\rm CP}_{\eta'
K^\pm}=0.02\pm0.04\,.
\end{equation}
Nevertheless, there is significant uncertainty in this constraint,
especially if $r_\phi\neq r_{\eta'}$. In the future, if the bounds on
the direct $CP$-asymmetries improve they will limit the size of the
strong phases compatible with RH new physics. This is in contrast to
the case of LH contributions where $r$ generically need not be as big,
reducing the sensitivity to measurements of direct $CP$-violation.

Depending on the specific operators involved, there are many other
constraints and correlations that can be studied. For instance, if the
chromomagnetic dipole operator is the dominant source of new physics,
there will also likely be a large contribution to the electromagnetic
dipole operator responsible for the decay $b\rightarrow
s\gamma$. However, in this case it has been shown that there can be a
significant contribution to $\sphi$ without violating the
$b\rightarrow s\gamma$
constraint~\cite{Kane:2002sp,Harnik:2002vs,Ciuchini:2002uv}. Another
stringent constraint comes from the limit on the
chromoelectric dipole moment of the strange quark derived from
experimental limits on the EDMs of the neutron and
$^{199}$Hg~\cite{Hisano:2003iw,Hisano:2004tf,Kane:2004ku}. Other
constraints like $b\rightarrow s\ell^+ \ell^-$ are starting to become
important as well. Of course, for any specific framework all available
constraints need to be studied, but such model dependent issues are
beyond the scope of this work.

\section{Conclusions}
\label{sec:conclusion}

We have studied a generic framework in which new physics contributions
only induce right-handed operators with a single CP violating
phase. This assumption leads to the following interesting
consequences:
\begin{itemize}
\item[(i)] Despite the opposite parity of the $\phi K$ and $\eta' K$
final states, RH operators can still account for the experimental data
thanks to the two-fold ambiguity in $S_{\phi,\eta'} =
\sin(2\beta+\Sigma_{\phi,\eta'})$.
\item[(ii)] The favored parameter region is more tightly constrained
for RH operators than for LH operators, and sizable new physics
contributions comparable in size to the SM are required.
\item[(iii)] The ease with which LH operators can accommodate $\sphi$
and $\seta$ is diminished when there are significant differences
between relative size of new physics contributions to the two modes.
\item[(iv)] Fitting the data with RH operators efficiently constrains
the new physics phase. Thus we find that the CP-asymmetries in
$B{\to}\phi,\eta' K_S$ are correlated with observables from the
$B_s-\dtlbar B_s$ system such as $\Delta m_s$ and the CP asymmetry in
$B_s\to \psi\phi$.
\end{itemize}
Work is underway to test the viability of this scenario in the
framework of a specific model of RH squark mixing, where the
constraints such as $b\rightarrow s\gamma$, $b\to sl^+ l^-$ and the
EDM of $^{199}$Hg can also be imposed. We conclude that it is too
early to rule out the contributions from right-handed operators to the
CP-asymmetries in $b\rightarrow s$ transitions.

\acknowledgments{GP thanks K.~Agashe, Y.~Grossman, M.~Neubert and
M.~Papucci for helpful discussions. This work was supported in part by
the DOE under contracts DE-AC03-76SF00098 and in
part by NSF grant PHY-0098840.}


\end{document}